\documentclass{aa}

\usepackage{astron}
\usepackage{graphics}
\usepackage{epsfig}

\begin {document}

\thesaurus{11.04.1  } 

\titlerunning{Anomalous SBF in NGC 4489}

\title{Anomalous Surface Brightness Fluctuations in NGC
4489\thanks{Based on observations performed at the European Southern
Observatory, La Silla, Chile; ESO program N$^o$ 62.N-0876.}}

\author{S.Mei \inst{1} \inst{2}, D.R. Silva \inst{2} \thanks{Visiting
Astronomer, Kitt Peak National Observatory, National Optical
Astronomical Observatories, operated by the Association of
Universities for Research in Astronomy, Inc., under cooperative
agreement with the National Science Foundation.}, P.J. Quinn \inst{2}}

\institute{ Observatoire Midi--Pyr\'en\'ees, 14 ave. E. Belin, 31400
Toulouse, France; mei@ast.obs-mip.fr \and European Southern
Observatory, Karl-Schwarzschild-Strasse 2, 85748 Garching, Germany }

\date{Received February 18, 2000; }

\maketitle

\begin{abstract}

Anomalously high K-band surface brightness fluctuations (SBF) have
been reported in NGC 4489 by Pahre \& Mould (1994), Jensen et al.
(1996) and Jensen et al.  (1998).  However, these conclusions were
uncertain because of relatively low signal-to-noise data.  New high
signal-to-noise data for NGC 4489 have been obtained at the NOAO/KPNO
2.1m and the ESO/La Silla 3.5m NTT telescopes.  Adopting the I-band
SBF distance modulus determined by Tonry et al.  (1990) and the
$\overline M_I$ versus (V--I) calibration of Tonry et al.  (2000), a
value of $\overline M_K =-6.18\pm0.09 $ mag was derived.  Relative to
the average empirical $\overline M_K$ derived for giant ellipticals by
Jensen et al.  (1998) ($\overline M_K=-5.61\pm0.12$), the detection of
an anomalous K-band SBF in NGC 4489 is confirmed at the two sigma
level.  Such anomalous fluctuations could be caused by an extended
giant branch, consisting of either intermediate-age AGB stars above
the tip of the first-ascent giant branch or high-metallicity
first-ascent giants, or by lack of a full understanding of the K--band SBF calibration. This result raises questions about the accuracy
of K-band SBF distance measurements for elliptical galaxies with
unknown stellar composition and underscores the need for $\overline
M_K$ measurements over a larger range of color and luminosity.


\end{abstract}

\section{ Introduction}

The Poissonian statistics of stellar populations in each pixel of a
galaxy image produces fluctuations. While the mean flux per pixel does
not depend on distance, the variance of these fluctuations is
inversely proportional to the square of the galaxy distance. When
properly calibrated this variance can then be used as a distance
indicator.  Surface Brightness Fluctuations (SBF) are defined as the
variance of the flux of the fluctuations normalized to the mean flux
of the galaxy in each pixel \cite{ts88}. The stars that contribute the
most to the SBF are then the brightest stars in the intrinsic
luminosity function, typically red giants in old stellar populations
(a recent review on the method has been given by Blakeslee et
al. 1999).

\noindent I--band SBF have being used successfully to measure
elliptical and S0 distances up to 4000 km s$^{-1}$ from ground based
telescopes and 7000 km s$^{-1}$ on the Hubble Space Telescope (HST)
\cite{sod95,sod96,aj97,ton97,tho97,lau98,pah99,bla99,ton00}.  Recently
SBF measurements have been attempted in the infrared K--band for
several reasons \cite{lup93,pah94,jen96,jen98,jen99}.

\noindent First of all the contrast between brightest red giants and
the underlying stellar populations is more extreme in the K--band,
producing larger SBF at any given distance. Second, at ground
facilities, K--band seeing is intrinsically better than I--band seeing
which also enhances the SBF contrast. Moreover the color contrast
between K--band SBF and external sources (globular clusters and
background galaxies) is larger than in the I--band.  Finally both
empirical and theoretical studies suggest that absolute K--band SBF is
almost constant with age for solar-metallicity stellar populations
\cite{pah94,jen96,jen98,jen99,mei99,liu00,mei00}.

\noindent These arguments suggest that in the infrared it should be
possible to determine SBF distances for galaxies at larger distances
than at optical wavelengths \cite{jen98}.  However, the K-band has one
major disadvantage for ground-based studies - the background is
significantly brighter making it much more difficult to detect and
correct for globular cluster and background galaxy contamination and
increasing the background shot noise.


\begin{figure*}

\centerline{\psfig{file=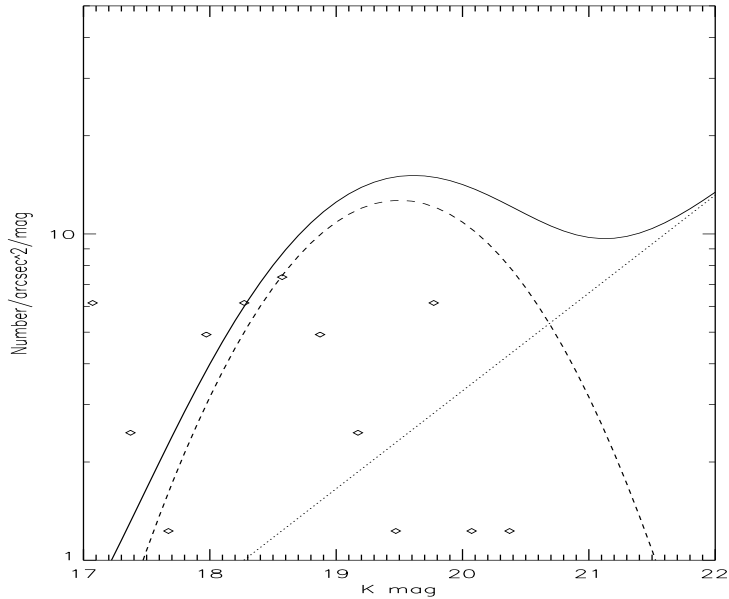,width=10cm,height=8cm}}

\caption{NGC 3379 external source luminosity function derived from the NTT
data.  The continuous line shows the sum of the adopted globular 
cluster plus background galaxy luminosity functions.  The dashed line 
shows the globular cluster luminosity function, while the dotted line 
shows the background galaxy luminosity function.  In total, 38
sources were detected between $\approx$ 15$\arcsec$ and $\approx$ 
60$\arcsec$.  } \label{fig-3379gc}

\end{figure*}


\noindent Initial K-band SBF studies suggested that the $\overline
M_K$ varied very little from galaxy-to-galaxy \cite{pah94,jen96}.  
However, these studies also argued that two Virgo ellipticals (NGC 
4365 and NGC 4489) had significantly brighter than average $\overline 
M_K$ values suggesting the presence of an extended giant branch.  This 
conclusion, however, was weakened by two problems: (1) no correction 
was made for globular cluster and background galaxy contamination; and 
(2) the SBF measurements were based on relatively low signal-to-noise 
data.  Jensen et al.  (1998) re-measured $\overline M_K$ for NGC 4365, 
using I-band data to determine a correction for globular cluster 
contamination, and concluded that NGC 4365 was a member of the 
background W cloud and the $\overline M_K$ for this galaxy was 
consistent with other Virgo ellipticals.  In short, NGC 4365 appears 
to have a normal giant branch.  No additional study of NGC 4489 has 
been published until now.

\noindent In this study, new observations with
significantly higher signal-to-noise and a new SBF analysis for NGC 
4489 are presented.  The observations are described in $\S$ 2, the SBF 
analysis in $\S$ 3 .  The results of this analysis are presented in 
$\S$ 4 and $\S$ 5.

\section{Observations}

\subsection{NGC 3379 and NGC 4489}

The galaxies NGC 3379 in the Leo cluster and NGC 4489 in Virgo were 
observed.  NGC 3379 has very well--established near--IR photometry
\cite{fro78,sil98} and has been already deeply studied in the I--band
by Sodemann \& Thomsen (1995).  It has been used here as a fiducial 
galaxy.  NGC 4489 is a low luminosity dwarf elliptical galaxy.  Its 
position angle and ellipticity vary in a significant way with radius.  
Its profile deviates from a De Vaucouleurs law, rising steeply towards 
the center.  It has a Mg$_{2}$ index of 0.198, different from the 
other Virgo galaxies.  

\begin{figure*}

\centerline{\psfig{file=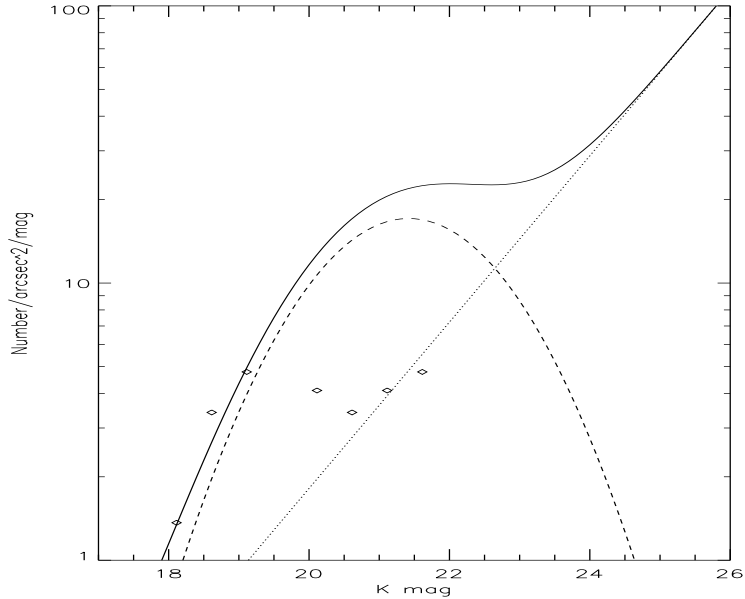,width=10cm,height=8cm}}

\caption{NGC 4489 external source luminosity function derived from the 
NTT data.  The continuous line shows the sum of the adopted globular 
cluster plus background galaxy luminosity functions.  The dashed line 
shows the globular cluster luminosity function, while the dotted line 
shows the background galaxy luminosity function.  In total, 62 sources 
were detected between $\approx$ 2$\arcsec$ and $\approx$ 60$\arcsec$.} 
\label{fig-4489gc}

\end{figure*}


\begin{figure*}[!htf]

\centerline{\psfig{file=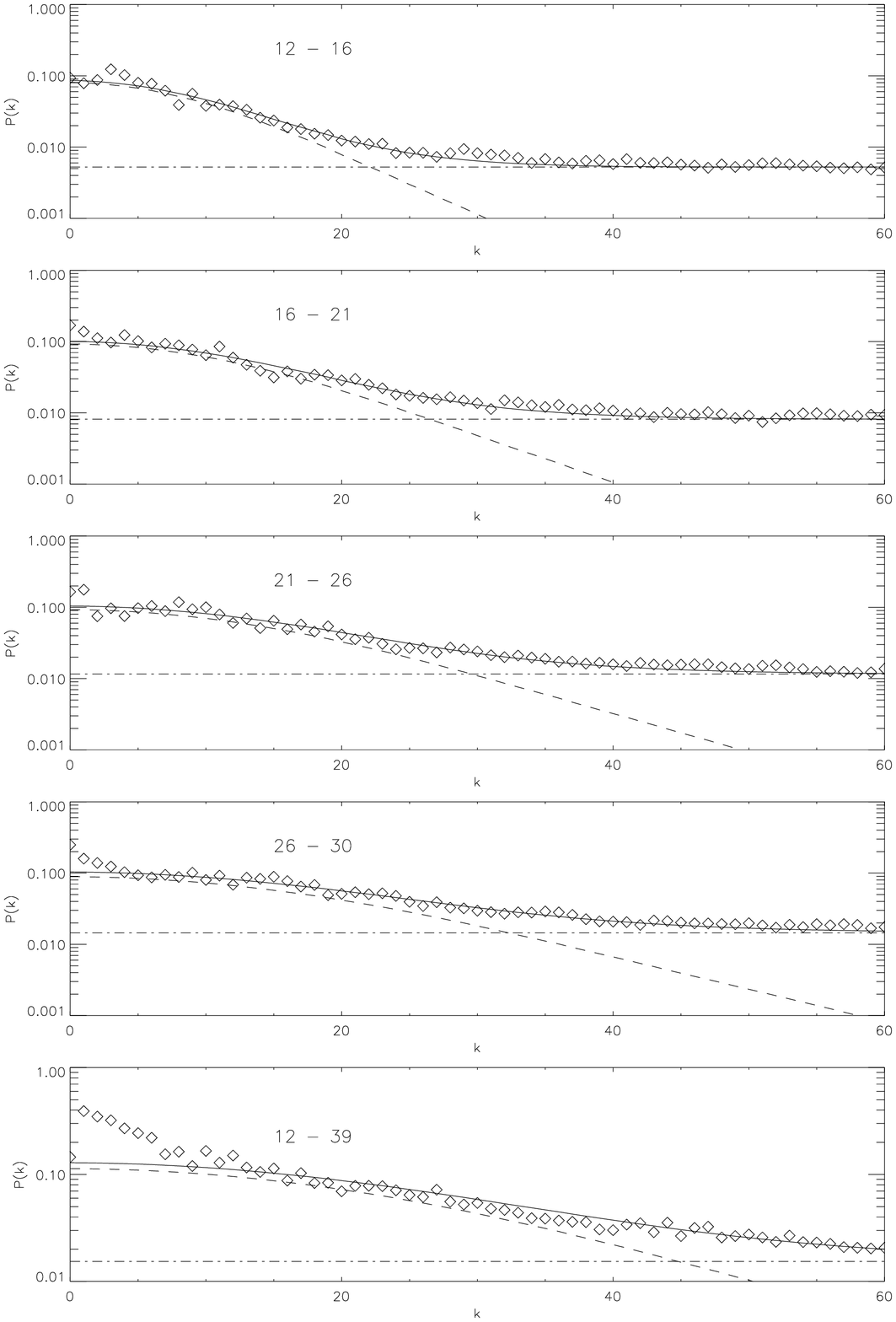,width=7cm,height=20cm}}

\caption{NGC 3379 NTT power spectra. The NGC 3379 power spectra as a
function of the wavenumber $k$ are shown for different annuli.  The
fit of the power spectrum is given by the continuous line, the PSF
power spectrum by the dotted line and the dashed line is the fitted
constant white noise spectrum.  See
Table~\ref{tab-ntt3379}. }\label{fig-an3379}

\end{figure*}

\subsection{Observations}

\noindent Data for NGC 3379 and NGC 4489 were obtained during two 
different observing runs.  The first dataset was obtained at the Kitt 
Peak National Observatory 2.1m on 1999 February 7.  The imaging camera 
was ONIS (OSU-NOAO Infrared Imaging Spectrometer) with a InSb 512 x 
1024 array detector.  The gain is 4 e$^{-}$/ADU, the read out noise 
$\approx$ 9 ADU, the dark current $\approx$ 1.2 e$^{-}/s$ .  The pixel 
scale was 0.34$\arcsec$ and the field of view 6$\arcmin$ x 3$\arcmin$.  
The filter that was used was a K short filter K$^{sh}$, centered at 
2.16 $\mu m$.  The K$^{sh}$ filter is centered at shorter wavelength 
than the standard K filter to reduce thermal background and at larger 
wavelength than the K$^{'}$ filter (Wainscoat and Cowie 1992).  Pahre 
and Mould (1994) estimated the difference between this filter and the 
standard K-band is of the order of 0.02 magnitudes for elliptical 
galaxies.  We do not scale our results to the standard K-band filter, 
meaning the calculated K-band SBF magnitudes are magnitudes in the 
K$^{sh}$ filter.

\noindent The seeing was 1$\arcsec$ for NGC 3379 and 0.9$\arcsec$ for
NGC 4489. The photometric zero-point and the extinction coefficient
were extracted from the observations of NICMOS red standards
\cite{per98}. A photometric zero-point $m_1=21.32$ mag and
an extinction coefficient of 0.1 mag in the K$^{sh}$ were determined.  
The sky brightness was on average 12.7 mag arcsec$^{-2}$.  NGC 3379 
was observed during a non-photometric period.

\noindent Each galaxy was observed in a sequence of
background-galaxy-background sets.  At each position, the integration
time was 10 x 6 secs.  After each set, a small random dither was
executed to facilitate removal of bad pixels.  The total exposure
times were 360 secs for NGC 3379 and 1700 secs for NGC 4489.

\noindent The second dataset was obtained using the 3.5 m New
Technology Telescope (NTT) at the European Southern Observatory, La 
Silla, Chile, on 1999 March 26.  The instrument used was SOFI (Son OF 
ISAAC) with a Hawaii HgCdTe 1024 x 1024 detector array.  The gain is 
5.53 e$^{-}/ADU$, the read out noise 2.1 ADU, the dark current $< 0.1 
e^{-}/s$ .  The pixel scale was 0.145$\arcsec$/pixel, with a field of 
view 2.47$\arcmin$ x 2.47$\arcmin$.  The filter was again a K$^{sh}$ 
filter.  The seeing was 0.75$\arcsec$ for NGC 3379 and 0.85$\arcsec$ 
for NGC 4489.  A photometric zero-point $m_1=22.32$ mag and an 
extinction coefficient of 0.07 mag were determined from NICMOS 
standards with solar-type colors \cite{per98}.  The sky brightness was 
on average 12.6 mag arcsec$^{-2}$.  Both galaxies were observed under 
photometric conditions.

\noindent NGC 3379 was observed in a sequence of galaxy-background
sets.  At each position, the integration time was 4 x 15 secs.  NGC
4489 was observed in a sequence of background-galaxy-background sets.
At each position, the integration time was 3 x 20 secs.  After each
set, a small random dither was executed to facilitate removal of bad
pixels.  The total exposure times were 1380 secs for NGC 3379 and 3780
secs for NGC 4489.

\noindent The two sets of data were similarly processed. Each 
galaxy image was background subtracted and then divided by a normalized 
dome flat field.  Dark current was subtracted by this operation.  Bad 
pixels and cosmic rays were eliminated by a sigma clipping algorithm 
while combining the images using the IRAF task IMCOMBINE. Sub-pixel 
registration was not used to avoid the introduction of correlated 
noise between the pixels into the images.

\begin{table*}
\caption{SBF measurements for various annuli of NGC 3379 from NTT 
observations} \label{tab-ntt3379}
\begin{flushleft}
\begin{tabular} {|c|c|c|c|c|c|c|c|} \hline 

Annulus  & P$_0$ & $\sigma_{P_0}$ & P$_0$/P$_1$ &$P_{es}$& $\overline{m}_{K}$ & $\sigma_{\overline{m}_{K}}$&$m_{cut}$\\ \hline
($\arcsec$)& ($ADU \ s^{-1}$) &($ADU \ s^{-1}$) && ($ADU \ s^{-1}$) &(mag)&(mag)&(mag) \\ \hline 

\hline

12 - 16 & 0.085 &0.007 & 25&0.0076&24.86&0.1& 19\\ \hline
16 - 21 & 0.096 &0.005 & 20&0.0076&24.71&0.07&19 \\ \hline
21 - 26 & 0.094 &0.003 & 15&0.0076&24.69&0.04& 19\\ \hline
26 - 30 & 0.100 &0.002 & 13&0.0076&24.66&0.04& 19\\ \hline
30 - 35 & 0.100 &0.001 & 9&0.0076&24.66&0.03& 19\\ \hline
35 - 39 & 0.094&0.002 & 8&0.0076&24.73&0.04& 19\\ \hline
2 - 39 & 0.093 &0.003 & 13.5&0.0076&24.70&0.03&19 \\ \hline

 \hline
 Mean  &  & & --&--&24.72&0.06&-- \\ \hline

\end{tabular}

\end{flushleft}

\end{table*}

\section{Surface Brightness Fluctuations Analysis}

Both datasets were analyzed by the SBF extraction standard technique 
used e.g. by Tonry and Schneider (1988).  A smooth galaxy model was 
built by fitting the galaxy isophotes and then subtracted from the 
image.  Visible external sources were not included in the fitting 
procedure.  External point sources were then identified using 
Sextractor \cite{ber96} and subtracted.  The residual image was 
smoothed on a scale ten times the width of the PSF. This smoothed 
image was subtracted from the residual image to correct residual 
sky-subtraction errors.  That image was divided by the square root of 
the galaxy model to make the amplitude of the SBF fluctuations 
constant across the entire image.  Finally, a point spread function 
(PSF) profile was determined from the bright stars in the image and 
normalized at 1 ADUs$^{-1}$.

\noindent Various annuli were then analysed.  The
photometric completeness function was calculated in each annulus by 
adding simulated point source images to the original, galaxy 
subtracted image.  The photometric completeness was calculated in 
intervals of 0.1 mag, as the ratio of the simulated and the extracted 
point sources.  The external point sources were masked up to a cut-off 
magnitude $m_{cut}$, whose value for each annulus is given in 
Table~\ref{tab-kp4489}, for KPNO observations, and in 
Table~\ref{tab-ntt3379} and Table~\ref{tab-ntt4489} for NTT 
observations.  Next, the image power spectrum was calculated in each 
annulus and normalized to the number of non-zero points in the 
annulus.  The power spectrum was azimuthally averaged.  

\noindent The total image power spectrum is the sum of two components: 
a constant power spectrum due to the white noise, $P_1$ and a power 
spectrum of the fluctuation and point sources.  Both components are 
convolved by the PSF in the spatial domain.  In the Fourier domain 
this second component is given by a constant $P_0$ multiplied by the 
power spectrum of the PSF:

\begin{equation}
E_{gal}= P_0 \  E_{PSF} +P_1 .
\end{equation}

\noindent To calculate $P_0$ and $P_1$, a robust linear least square 
fit, minimizing absolute deviation (Numerical Recipes, Press et al.  
1992), was made to the power spectra of each annuli.  Low (typically 
less than 4) wave number points were excluded from the fit since they 
are contaminated by the galaxy subtraction and subsequent smoothing 
errors, plus residual background variance contribution, as pointed out 
e.g. by Jensen et al.  (1999).  

\noindent $P_0$ was then corrected for the contribution of point 
source fluctuations $P_{es}$, estimated from the
equations:

\begin{equation}
P_{es}=\sigma^2_{gc}+\sigma^2_{bg}
\end{equation}

\noindent as described by Blakeslee \& Tonry (1995): $\sigma^2_{gc}$
is the contribution to the fluctuations given by globular clusters,
$\sigma^2_{bg}$ is the contributions by background galaxies.  It as
assumed that the luminosity function of the globular clusters is given
by:

\begin{equation}
N_{gc}(m) = \frac{N_{ogc}}{\sqrt{2\pi}\sigma} 
e^{\frac{-(m-m_{peak})^2}{2\sigma^2}},
\end{equation}

\noindent with $\sigma = 1.35$, $M_{peakV}=-7.5$ \cite{fer00} and
(V--K) = 2.23 for NGC~4489 and $m_{peakB}= 22.70$ and $\sigma = 0.9$ from Pritcher \& Van den Bergh (1985) for NGC~3379 and standard (V--K) color ($\approx$ 3) from Kissler--Patig (2000).  We have verified that 
exact details in the adopted globular cluster and background galaxy 
luminosity functions have little effect on the final measurement of 
SBF amplitudes, as pointed out i.e. from Blakeslee et al.  (1999a).  
For the background galaxies a power-law luminosity function was 
assumed:

\begin{equation}
N_{bg}(m) = N_{obg} 10^{\gamma m } 
\end{equation}

\noindent with $\gamma = 0.3$ \cite{cow94}.  $N_{ogc}$
and $N_{obg}$ were estimated by the fitting these equations to the 
observed composite luminosity function determined from the external 
sources extracted from the image.  The actual $m_{cut}$ used are given in
Table~\ref{tab-ntt3379}, Table~\ref{tab-kp4489}, and 
Table~\ref{tab-ntt4489}, while the observed external source luminosity 
functions for NGC 3379 and NGC 4489 are shown respectively in 
Figure~\ref{fig-3379gc} and Figure~\ref{fig-4489gc}, as derived from 
the NTT dataset.  The fit has been made as from the sum of the 
globular cluster plus background galaxies luminosity function:

\begin{eqnarray}
 N(m) =&N_{gc}(m) + N_{bg}(m) \\ \nonumber
&=\frac{N_{ogc}}{\sqrt{2\pi}\sigma} \ e^{\frac{-(m-m_{peak})^2}{2\sigma^2}} +  N_{obg} \ 10^{\gamma m}.
\end{eqnarray}

\noindent We keep all the parameters fixed in the fit, but $N_{ogc}$
and $N_{obg}$.  We have assumed a galaxy distance modulus of 30.04 mag
for NGC 3379 and of 31.15 mag for NGC 4489 (see discussion in next
section).  Identified foreground stars were not included in the fit.

\noindent $P_{es}$ was then calculated as the sum of:

\begin{eqnarray}
\sigma^2_{gc}=&\frac{1}{2} N_{ogc} 10^{0.8[m_1-m_{peak}+0.4\sigma^2ln(10)]}\\ \nonumber
& erfc[\frac{m_{cut}-m_{peak}+0.8\sigma^2 ln(10)}{\sqrt{2}\sigma}]
\end{eqnarray}

\noindent and

\begin{equation}
\sigma^2_{bg}=\frac{N_{obg}}{(0.8-\gamma) ln(10)}10^{0.8(m_1-m_{cut})+\gamma(m_{cut})}.   
\end{equation}

\noindent where $m_{1}$ is the zero magnitude which corresponds to a
flux of 1 ADUs$^{-1}$.

\noindent Finally, the apparent SBF magnitude was computed as:

\begin{equation}
\overline{m}_{K} = -2.5 log(P_0-P_{es}) + m_{1} -\epsilon_{ext} sec(z)
\end{equation}

\noindent where $\epsilon_{ext}$ is the extinction coefficient, and
sec($z$) the airmass for the observations.  Color term and redshift
corrections are negligible \cite{liu00}.  No galactic foreground
reddening corrections were applied given that E(B--V) = 0.028 for
NGC~4489 and E(B--V) = 0.028 for NGC~3379, implying $A_K$ = 0.01 for
NGC~4489 and 0.009 for NGC~3379 \cite{sch98}.  We measured
$\epsilon_{ext}=0.1$ and $m_{1}=21.21 \pm 0.03$ mag for the Kitt Peak
data and $\epsilon_{ext}=0.07$ and $m_{1}=22.18 \pm 0.03$ mag for the
NTT data.  The error on $\overline{m}_{K}$ is given as the standard
deviation of the different annuli considered plus the entire field.
To this error, the errors due to the zero point magnitude and
$\epsilon_{ext}$ calibration have been added in quadrature.  When a
distance estimation from $\overline {m}_{K}$ is made, the uncertainty
of the SBF absolute magnitude calibration $\overline {M}_{K}$
calculated from previous K-band SBF observations \cite{jen98} is also
added in quadrature.


\begin{figure*}[!ht]

\centerline{\psfig{file=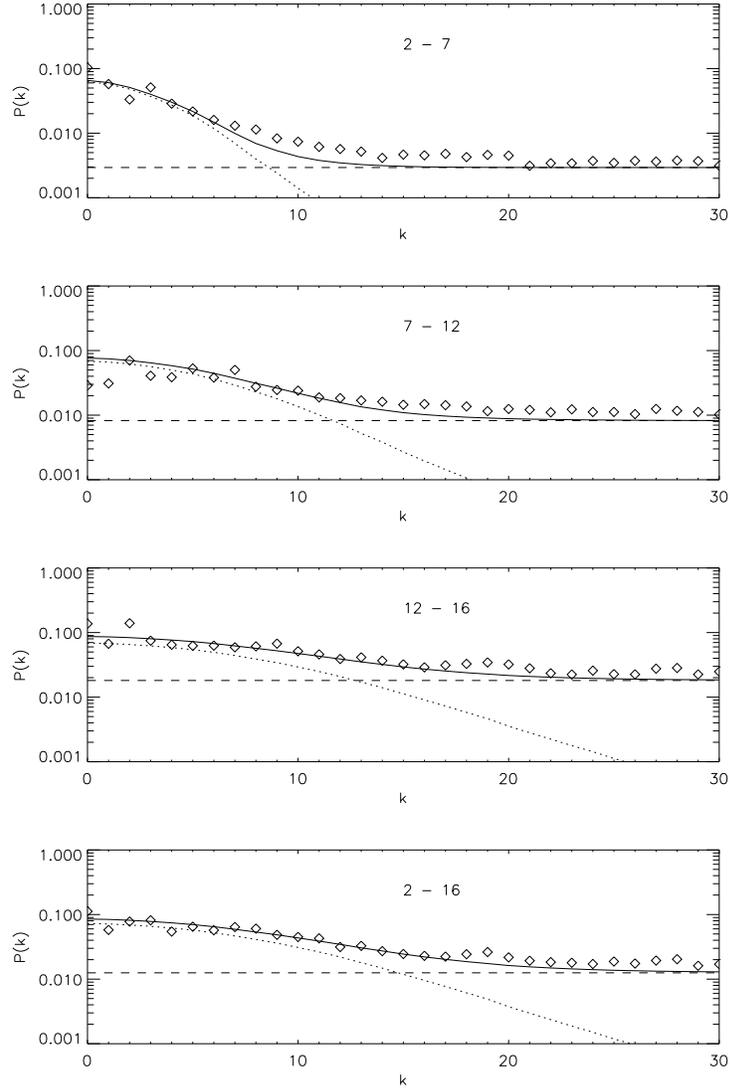,width=10cm,height=15cm}}

\caption{NGC 4489 NTT power spectra. The NGC 4489 power spectra are
shown as a function of the wavenumber $k$ from NTT observation in
various annuli.  The fit of the power spectrum is given by the
continuous line, the PSF power spectrum by the dotted line and the
dashed line is the fitted constant white noise spectrum.  See
Table~\ref{tab-ntt4489}.}
\label{fig-an4489}

\end{figure*}

\section{Results}
\subsection{NGC 3379}

The resultant NGC 3379 power spectra derived from the NTT data are 
shown in Figure~\ref{fig-an3379}.  The results for each annulus for 
NTT data are listed in Table~\ref{tab-ntt3379}.  The standard 
deviation from each annulus calculated as the variance among different 
low wavelength cuts is listed.  To the fitting error, the errors due 
the photometric zero point calibration and to external source residual 
contribution subtraction were added in quadrature.  The Kitt Peak data 
for NGC 3379 were obtained under non-photometric conditions and are 
excluded from the final analysis.  In both NGC 3379 datasets, here is 
a tendency to have fainter fluctuations in the center of the galaxy
\cite{sod95}.

\noindent From NTT data a value of $\overline{m}_{K} = 24.72
\pm 0.06$ mag was derived.  The error is given by adding in quadrature
the standard deviation of the magnitudes derived in each annulus 
divided by the number of considered values.  
 Pahre and Mould (1994) derived a value of $\overline{m}_{K} = 24.48 \pm 0.1$ mag for NGC 3379 from the 
analysis of annuli between $\approx 4\arcsec$ and $\approx 45
\arcsec$. They did not correct for contamination by external source
contribution correction.  This could explain why they reported 
slightly larger fluctuations.  Based on the Tonry et al.  (1990) 
measurements of $\overline{m}_I= 28.63 \pm 0.04$ mag and the most recent $(V-I) = 1.223 \pm 0.015$  mag \cite{fer00} and the Tonry et al.  (2000) calibration:

\begin{equation}
\overline{M}_I=(-1.74 \pm 0.08)+(4.5 \pm 0.25) \ ((V-I)-1.15), 
\end{equation}

\noindent a distance modulus of ($\overline{m}_{I}$ - $\overline{M}_{I}$)=30.04 $\pm$ 0.11 mag was adopted.

\noindent Using the adopted I-band distance modulus, an absolute
magnitude of $\overline{M}_{K} =-5.32 \pm 0.13$ mag was derived.  This
value is consistent with measurements for other ellipticals presented
in Jensen et al.  (1998) sample and with nearly solar metallicity, 15
Gyr old models from Bruzual \& Charlot (2000), as analyzed by Mei et
al.  (1999), Mei et al.  (2000) and Liu et al.  (2000) and from
Worthey, as analyzed by Jensen et al.  (1998).


\begin{table*}[!ht]
\caption{SBF measurements for various annuli of NGC 4489 from KPNO
observations} \label{tab-kp4489}
\begin{flushleft}
\begin{tabular} {|c|c|c|c|c|c|c|c|c|} \hline 

Annulus & P$_0$ & $\sigma_{P_0}$ & P$_0$/P$_1$&$P_{es}$ & $\overline{m}_{K}$ & $\sigma_{\overline{m}_{K}}$&$m_{cut}$\\ \hline
($\arcsec$)& ($ADU \ s^{-1}$) &($ADU \ s^{-1}$) && ($ADU \ s^{-1}$) &(mag)&(mag)&(mag) \\ \hline 

\hline
2 - 7 & 0.032&0.005 & 10&0.0009&24.85&0.15&19.9 \\ \hline
7 - 12 & 0.026 &0.004 & 4&0.0009&25.08&0.14&19.9 \\ \hline
12 - 16 & 0.030 &0.005 & 3&0.0009&24.92&0.13&19.9 \\ \hline
2 - 16&0.026&0.001&3&0.0009&25.08&0.03&19.9\\ \hline

 \hline
 Mean  &  & & --&--&24.98&0.06&-- \\ \hline

\end{tabular}

\end{flushleft}

\end{table*}



\begin{table*}
\caption{SBF measurements for various annuli of NGC 4489 from NTT
observations} \label{tab-ntt4489}
\begin{flushleft}
\begin{tabular} {|c|c|c|c|c|c|c|c|c|} \hline 

Annulus  & P$_0$ & $\sigma_{P_0}$ & P$_0$/P$_1$  &$P_{es}$ &$\overline{m}_{K}$ & $\sigma_{\overline{m}_{K}}$&$m_{cut}$\\ \hline
($\arcsec$)& ($ADU \ s^{-1}$) &($ADU \ s^{-1}$) && ($ADU \ s^{-1}$) &(mag)&(mag)&(mag) \\ \hline 

\hline
2 - 7 & 0.074&0.01 & 50&0.0022&24.96&0.18&19.9 \\ \hline
7 - 12 & 0.068 &0.007 & 40&0.0022&25.04&0.11&19.9 \\ \hline
12 - 16 & 0.076 &0.008 & 25&0.0022&24.91&0.13&19.9 \\ \hline
2 - 16&0.070&0.002&25&0.0022&25.00&0.03&19.9\\ \hline

 \hline
 Mean  &  & & --&--&24.97&0.06 &--\\ \hline

\end{tabular}

\end{flushleft}

\end{table*}


\subsection[NGC 4489]{NGC 4489}

The power spectra for NGC 4489 as derived from the NTT data are shown 
in Figure~\ref{fig-an4489}.  The KPNO data produce similar results, 
albeit with lower signal-to-noise.  External source detections from 
the NTT data were used to mask the external sources from the KPNO 
dataset after appropriate shifting and scaling.  The results for each 
annulus for the respective datasets are listed in 
Table~\ref{tab-kp4489} and Table~\ref{tab-ntt4489}.  The standard 
deviation from each annulus calculated as the variance among different 
low wavelength cuts is listed.  To the fitting error the errors due to
the photometric zero point calibration and to external source residual 
contribution subtraction are added in quadrature.  

\noindent The signal-to-noise
ratio P$_0$/P$_1$ for NTT data in the four annuli permit us to assess 
the SBF absolute magnitude for this galaxy with a percent error of 
approximately $ 10\%$ (Mei et al.  2000).  From the NTT data we derive 
a value of $\overline{m}_{K} = 24.97 \pm 0.06$ mag and from the KPNO 
data $\overline{m}_{K} = 24.98 \pm 0.06$ mag.  The errors were 
calculated by adding in quadrature the standard deviation of the 
magnitudes derived in each annulus, divided by the 
number of considered values.  Based on the most recent $\overline{m}_{I} =
29.10 \pm 0.13$ mag and $(V-I) = 1.081 \pm 0.015$ \cite{fer00}, and using the
$\overline{M}_{I}$  vs (V--I) Tonry calibration from Tonry et al.  (2000), a distance 
modulus for this galaxy equal to ($\overline{m}_{I}$ - 
$\overline{M}_{I}$)=31.15 $\pm$ 0.11 mag was adopted.  K--band SBF 
absolute magnitudes of $\overline{M}^{NTT}_{K} = -6.18 \pm 0.13$ mag 
and $\overline{M}^{KPNO}_{K} = -6.17 \pm 0.13$ mag were determined.  
The average of these two values is $\overline{M}_{K} = -6.18 \pm 0.09$ 
mag.  

\noindent This result is in agreement with the Jensen et al.  (1998)
measurement of $\overline{m}_{K} = 24.99 \pm 0.35$ mag.  It is not
comparable with previous measurements by Pahre \& Mould (1994) and
Jensen et al.  (1996), because of the different treatment of the external source contribution to the fluctuations
between this study and the previous studies.

\noindent The observed $\overline{M}_{K}$ is inconsistent with the
mean empirical value derived by Jensen et al. (1998) ($-5.61 \pm
0.12$) at the two-sigma level, in the sense that NGC 4489 has brighter
K-band surface brightness fluctuations than seen in bright, nearby,
cluster ellipticals.  This would appear to confirm the suggestion by
Pahre \& Mould (1994) and Jensen et al. (1996) that NGC 4489 contains
bright stellar component above the tip of the first ascent giant
branch, a so-called extended giant branch population.  Those initial
studies argued that such a population could be similar to the K-band
bright stellar population reported by Freedman (1992) and Elston \&
Silva (1992) in M32 (see also Luppino \& Tonry 1993 and Davidge 2000).
The exact nature of such extended giant branches remains
controversial, with most studies arguing for the presence of an
intermediate-age ($\sim$ 2 Gyr) AGB.  The most likely alternative
population currently appears to be high-metallicity first ascent giant
stars, brighter than normally observed in Galactic globular clusters
(Guarnieri et al. 1997).

\noindent However, NGC 4489 is in fact significantly fainter and bluer
(V--I $=$ 1.081) than the mean elliptical in the Jensen et
al. calibration.  This suggests that a comparison between the
$\overline{M}_{K}$ measured for NGC 4489 and the Jensen et al. mean
value may not be strictly valid.  Since there are no published
empirical $\overline{M}_{K}$ measures for ellipticals for V--I $\sim$~1, only theoretical predictions can be used to probe this color range.
Unfortunately, currently available theoretical calculations of
$\overline{M}_{K}$ provide conflicting results.  Models based on the
latest Bruzual \& Charlot models (e.g. Liu et al. 2000; Mei et
al. 2000), predict $\overline{M}_{K} =$ 4.7 -- 5.5 for galaxies with
V--I color similar to NGC 4489 for single-burst old populations.  For these models, the near-IR SBF
magnitudes become brighter as the galaxies becomes redder.  On the
other hand, independent models by Blakeslee, Vazdekis, \& Ahjar (2000)
predict $\overline{M}_{K}$ as bright as --6.5 in this V--I color
range, with near-IR SBF magnitudes becoming somewhat fainter as
galaxies become redder.

\noindent In the end, one thing is certain: NGC 4489 illustrates that
the Jensen et al. empirical mean $\overline{M}_{K}$ is not universal.
Whether or not this means that NGC 4489 has an unusual stellar
population mixture cannot be answered without further empirical
observations of ellipticals with similar colors.






\section{Summary and Conclusions}

New, high signal-to-noise K-band SBF measurements have been obtained
for NGC 3379 and NGC 4489.  In summary:

\begin{itemize}

\item K--band SBF $\overline M_{K}$ values of $-5.32 \pm 0.13$ mag for
NGC 3379 and $-6.18 \pm 0.09$ mag for NGC 4489 have been derived.
Although the new apparent SBF magnitudes presented here are
independent observations and are more accurate for NGC 4489 than
previous determinations, the final K-band SBF absolute magnitudes rely
on the accuracy of the respective distance moduli determined in the
I-band.

\item Keeping in mind the caveat above about absolute magnitudes, the
measurement by Jensen et al.  (1998) and Pahre \& Mould (1994) of
large K--band SBF in NGC 4489 has been confirmed.  These fluctuations
could be the signature of an extended giant branch, but this cannot be
confirmed until K-band SBF magnitudes are determined for a larger
sample of galaxies similar in color to NGC 4489.

\end{itemize}

These results raise questions concerning the accuracy of K--band SBF
distance measurements for elliptical galaxies.  SBF measurements have
been extended to this band in the belief that one could determine with
the same accuracy K--band SBF distances to larger distances than
possible at optical wavelengths.  This is possible only if the
contribution of external sources to the observed image fluctuations
can be correctly estimated and if the K--band SBF absolute magnitude
can be precisely calibrated.  The result of this paper concerns the
issue of calibration.  The presently small sample of K--band
measurements (around 20) would seem to suggest that $\overline M_K$
varies very little from galaxy to galaxy.  However, our result -- a
galaxy with fluctuations different than the current average,
illustrates the importance of an extended K--band calibration over a
larger range of luminosity and color.  Only larger samples will allow
us to determine whether or not NGC 4489 (or for that matter M32) is an
exceptional case, and to understand the real potential of K--band SBF
distance determinations.

\begin{acknowledgements}

S. Mei is grateful to P. Rosati and M. Romaniello for the useful
discussions and acknowledges support from the European Southern
Observatory Studentship programme and Director General's Discretionary
Fund.

\end{acknowledgements}

\end{document}